\documentclass[aps,prl,twocolumn,showpacs,superscriptaddress,groupedaddress]{revtex4}  
\usepackage{graphicx}  
\usepackage{dcolumn}   
\usepackage{bm}        
\usepackage{amssymb}   

\begin{document}

\title{Capillary fracture of ultrasoft gels: heterogeneity and delayed nucleation}

\author{Marion Grzelka}
\affiliation{Ecole Normale Sup\'erieure de Cachan, 94235 Cachan, France}
\affiliation{Department of Physics, North Carolina State University, Raleigh, NC, USA}
\author{Joshua B. Bostwick}
\affiliation{Department of Mechanical Engineering, Clemson University, Clemson, SC, USA}
\author{Karen E. Daniels}
\affiliation{Department of Physics, North Carolina State University, Raleigh, NC, USA}

\date{22 Aug 2016}

\begin{abstract}
A droplet of surfactant spreading on an ultrasoft ($E \lesssim 100$ Pa)
gel substrate will produce capillary fractures at the gel
surface; these fractures originate
at the contact-line and propagate outwards in a starburst pattern.
There is an inherent variability in both the number of fractures formed and the
time delay before fractures form. In the regime where single fractures form, we
observe a Weibull-like distribution of delay times,
consistent with a thermally-activated process.
The shape parameter is close to 1 for softer gels (a Poisson process),
and larger for stiffer gels (indicative of aging).
For single fractures, the characteristic delay time is primarily set by the
elastocapillary length of the system, calculated from the differential in surface
tension between the droplet and the substrate, rather than the elastic modulus as for stiffer systems.
For multiple fractures, all fractures appear simultaneously and long delay times are suppressed.
The delay time distribution provides a new technique for probing
the energy landscape and fracture toughness of ultrasoft materials.
\end{abstract}

\maketitle 

The failure of soft materials is highly relevant to many biological and
medical processes such as cellular dynamics
\citep{Levental2006,Beaune2014}  or drug delivery
over mucus membranes \citep{Khanvilkar-2001-DTT, Haitsma-2001-ESD}.
These highly-deformable materials, which include gels, elastomers, and biological tissues, can have elastic moduli as low as $10-100$~Pa, and are
sufficiently soft that they
cannot support their own weight when freestanding.
Material strength comes from cross-linked polymers that are known to have heterogeneous mechanical properties \citep{Goldbart1989a, meyers2008}, which makes performing traditional materials tests challenging.
In this paper, we present a novel method for probing the strength of
ultrasoft materials on the millimeter scale by using the surface tension (capillarity)
of liquid droplets to provide well-controlled, but weak, surface forces.
Our technique draws on both prior experiments on delayed fracture \cite{Bonn-1998-DFI}
and recent advances in understanding the spreading, wetting,
and material failure in this elastocapillary regime \citep{Jerison2011,Das2011,Bostwick2013,bostwick2014elastocapillary}.

\begin{figure}
\includegraphics[width=\linewidth]{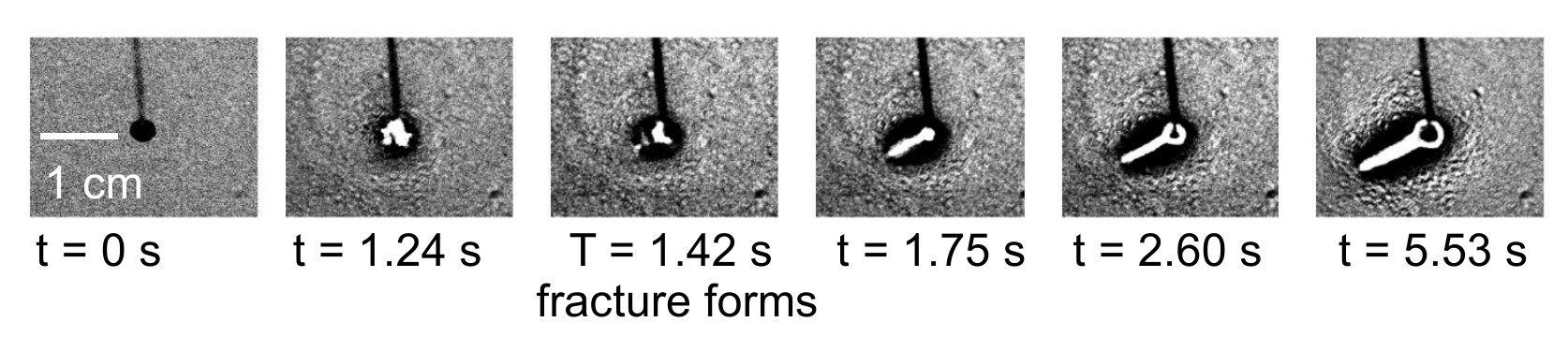}
\caption{Typical delayed fracture time evolution from Series 1 (Table~\ref{tab:seriespar}); a needle deposits a droplet at $t=0 s$ and a fracture with $n=1$ is nucleated at $t=1.42 s$ that propagates outwards from the contact-line.}
\label{fig:exp}
\end{figure}

In our experiments, we deposit a droplet of surfactant-water solution
on the surface
of an agar substrate and observe the formation of starburst-shaped
capillary fractures
that propagate radially outward from the contact-line. It has previously been observed that
the number of fractures formed is controlled by the ratio of the surface tension
contrast between the droplet and the gel $\Delta \sigma = \sigma_g- \sigma_d$
and the elastic modulus $E$ of the gel substrate \citep{Daniels2007}.
Similar instabilities have been observed on various gel/fluid combinations
\citep{Spandagos2012, Spandagos2012b, Foyart2013}.
Typical fracture experiments use increasing stress to find a fracture threshold or cyclic load to determine fatigue. In contrast, we apply a constant force in a technique similar to that of \cite{Bonn-1998-DFI}, where agarose gel rods ($E \approx 50$~kPa) were bent to
a fixed strain and held until material failure arose through a thermally-activated process. This method allows
for probing  the energetics of the crosslinks from the statistical
distribution of the delay times.  We measure histograms for the delay time and number of fractures, revealing that the nucleation process is thermally-activated; this method allows for estimating the typical size of energy barriers \cite{pomeau1992}.

It is helpful to contrast our approach with classic droplet-spreading
experiments on solid \citep{Tanner1979},
strong gel \citep{Szabo-2000-SLG,Kaneko-2005-KFS,Banaha2009,Kajiya2013} ($E =
75-150$~kPa) or liquid \citep{Hoult1972} substrates.
The elastocapillary length $\lambda = \sigma_d / E$
sets the scale of elastic deformation in problems involving the interactions
between liquids and compliant substrates.
(For reference, a droplet of water
($\sigma_d=72$~mN/m) wetting a glass substrate ($E=70$~GPa) produces
negligible deformations  of size $\lambda \sim 10^{-12}$~m.)
Recently, attention has shifted to soft substrates
\citep{Pericet-Camara2008,Kajiya2011a,Das2011a,Jerison2011,Style2012,Marchand2012,
Weijs2013,Henann2014,Park2014,Style2016,Andreotti2016a}, where these deformations are no longer negligible.
For example, \cite{Jerison2011} used fluorescence confocal microscopy to
quantify the deformations produced by a droplet of water on a silicone
gel substrate ($E\sim10$~kPa); these deformations are on the scale of $\lambda \sim
10^{-6}$~m.
The substrates we use in our experiments have an elastic modulus
$E \lesssim 100$~Pa with deformations $\lambda \sim 10^{-3}$~m, which is a
length scale on the same order as the droplet radius. Therefore, we refer to
these materials as {\itshape ultrasoft}, with the resulting
elastocapillary deformations large enough to cause the fracture of the
substrate. Understanding the various regimes
in which elastocapillary deformations are significant will aid in
understanding the physics of fracture for soft materials.

\paragraph*{Experiment:}
We investigate the fracture of ultrasoft gel substrates, composed of
agar (polysaccharide with galactose subunits) dissolved in deionized water. The
concentrations investigated range from $\phi = 0.115-0.127$~\%w agar, which is
above the gel transition at $\phi_c= 0.013\%$ at $20.0^\circ$C
\citep{Tokita1987}. These concentrations correspond to  
an elastic modulus $E=40-60$~Pa; values of $E$ are obtained from the shear modulus measurements of \cite{Tokita1987} under the  
assumption that the Poisson ratio $\nu = 1/2$ due to
the incompressibility of the water phase. Due to the strong dependence $E(\phi)$ and the aging of gels \citep{Scherer1988,Hodge1995}, we find that repeatability of experiments requires careful control of the preparation
process. Gels are prepared by dissolving agar powder into $25$~mL of deionized water at a temperature of $90^\circ$C. The solution is poured into individual Petri dishes (diameter $9.5$~cm) and cooled overnight at room temperature $20.5\pm0.5^\circ$C. The final thickness of each substrate was measured to be $h=3.0\pm 0.2$~mm.

\begin{table}
\begin{center}
\begin{tabular}{|c|c|c|c|c|c|c|c|}
\hline
       & $\phi$ &  $E$ & $\chi$ & $\sigma_d$ & $\delta=\frac{\Delta\sigma}{E}$ & $\lambda=\frac{\sigma_d}{E}$\\
Series &  [\%]    &  [Pa]  &    [ppm] &  [mN/m]      & [mm]          & [mm] \\ \hline
1 & 0.115& 41.1 & 80 & 61.2 & 0.19 & 1.49 \\ \hline
2 & 0.115 & 41.1 & 200 & 59 & 0.24 & 1.44 \\ \hline
3 & 0.123 & 52.5 & 250 & 58 & 0.21 & 1.10 \\ \hline
4 & 0.127 & 59.1 & 300 & 57 & 0.20 & 0.96\\ \hline
\end{tabular}
\caption{Experimental parameters for the four data series, with $\Delta \sigma$ calculated for $\sigma_g = 69$~mN/m for all gels.}
\label{tab:seriespar}
\end{center}
\end{table}

When a liquid droplet is placed on the surface of the gel, surface forces cause fractures to form, as shown in Fig.~\ref{fig:exp}. To control the magnitude of these forces, we utilize Triton X-305 surfactant (Dow Chemical, octylphenoxy polyethoxy ethanol) dissolved in deionized water at concentration $\chi$ ranging from $80-300$~ppm. The droplet surface tension $\sigma_d$ varies from $61.2-55$~mN/m with larger $\chi$ yielding smaller $\sigma_d$ \citep{Zhang2005}.  A volume-controlled syringe pump releases droplets of volume $V=21 \pm 0.1 \, \mu$L from a height $H=3.2$~cm directly above the center of 
the gel substrate.  For simplicity, we assume the surface tension of the gel $\sigma_g$ is constant and we observe that the wetting behavior is primarily controlled by the
surface tension contrast $\Delta\sigma \equiv \sigma_g-\sigma_d$. Note that the shape of the droplet (and hence the contact line radius and surface force) are also important factors; this consideration is discussed in more detail in
\cite{Bostwick2013}.

Fractures are visualized using shadowgraphy:
a point source of light passes through a converging lens resulting in parallel light that is transmitted through the sample, which is subject to refraction due to the variations in the index of refraction for the gel and the droplet. The image is captured on a ground glass screen located above the sample using a digital camera operating at frequency $f=15$~Hz. Our technique allows for the measurement of both the number of fractures $n$ and the delay time $T$ before fractures initiate. 
We calculate both the time $t=0$ when the droplet first contacts the substrate,  and the delay time $T$ when a fracture forms, via an ad hoc image-processing code that identifies changes in the standard deviation of the image light intensity.

In previous work, \cite{Daniels2007} observed significant variation in the number of fractures observed for a fixed set of experimental parameters ($\sigma_d$, $E$). In order to probe how such variation arises, as well as
the statistics of thermal activation, we minimize this variability. In addition to the strategies mentioned
above (correcting agar concentration during pouring, aging gels for a consistent time, and using a syringe pump to deposit droplets), we embed the entire apparatus in a sandbox to damp out the acoustic noise and building vibrations  that can  prematurely initiate fractures.
To obtain statistics to quantify these variations, we perform experiments on approximately $1200$ samples divided among the four series listed in Table~\ref{tab:seriespar}.
This range of values covers a regime in which starbursts with $n=0$ to $n=4$ fractures are formed.

\paragraph*{Results:}
It has been previously reported \citep{Daniels2007} that the mean number of fractures $\langle n \rangle$ increases as a function of the quantity
\begin{equation}
\delta \equiv \Delta \sigma/E.
\end{equation}
This quantity is related to the elastocapillary length $\lambda$ described above, since $\sigma_g$ is approximately constant. However, note that $\delta$ and $\lambda$ have the opposite trend as a function of $\sigma_d$.  We quantify our results using both $\delta$ and $\lambda$, and determine that $\delta$ is the more natural choice for these experiments.   For agar, $\sigma_g$ is just slightly less \citep{Daniels2007} than the value for pure water (72~mN/m) and is difficult to measure accurately. In the analyses that follow we use $\sigma_g = 69$~mN/m.

As illustrated in Fig.~\ref{fig:exp}, fractures do not necessarily nucleate immediately after the droplet is placed on the gel substrate, but after some delay $T$. We can understand this observation by considering the
elastic deformations within the substrate, induced by the wetting forces between the droplet and the gel \citep{Das2011a, Jerison2011, Bostwick2013, Bardall2016}. In general, the state of stress within the substrate is not quite large enough to cause material failure. Instead, the gel remains in this deformed elastic state until something triggers a failure, either an external perturbation or a thermal fluctuation.

\begin{figure}
\includegraphics[width=0.8\linewidth]{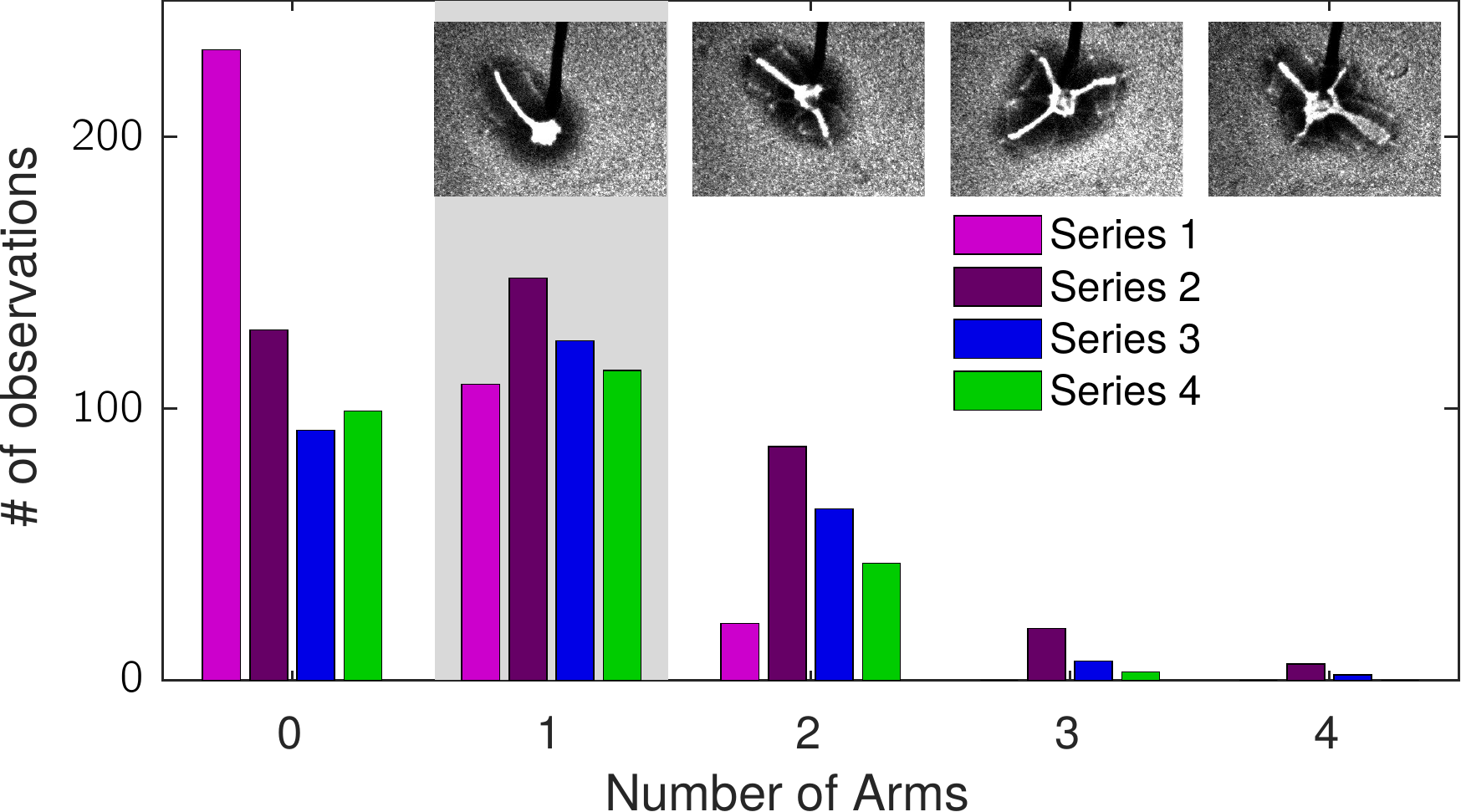}
\caption{Histogram of the number of fractures $n$ formed for each of the data series given in Table~\ref{tab:seriespar}. The set of runs with $n=1$ (gray rectangle) form the basis of our analyses. Series 2 is used to investigate multiarm statistics, and sample images are drawn from this Series.} \label{fig:fracnum}
\end{figure}

To quantify the dependence of the delay time $T$ on the properties of the gel substrate, we select three pairs of $(E, \sigma_d)$ values for which $n=1$  is highly likely (see Fig.~\ref{fig:fracnum}). (This was done empirically by selecting three values of $E$, and varying  $\sigma_d$ (and hence $\Delta\sigma$) until we observed  $n=1$ fractures in approximately $1/3$ of the trials.) These pairs correspond to approximately constant $\delta \approx 0.2$~mm, a length consistent with the size of observed surface deformations \citep{Daniels2007}. One  set of parameters (Series 1) provides a control series, by matching the value of $E$ for Series 2, while still having $n=1$ as a highly likely outcome.  Series 3 and 4 increase $E$ and decrease $\sigma_d$ in order to maintain an approximately consistent histogram ${\cal P}(n)$. Series 2 contains enough fractures with $n>1$ to allow for a semi-quantitative investigation of the delay statistics of multiarm starbursts.

\begin{figure}
\includegraphics[width=0.8\linewidth]{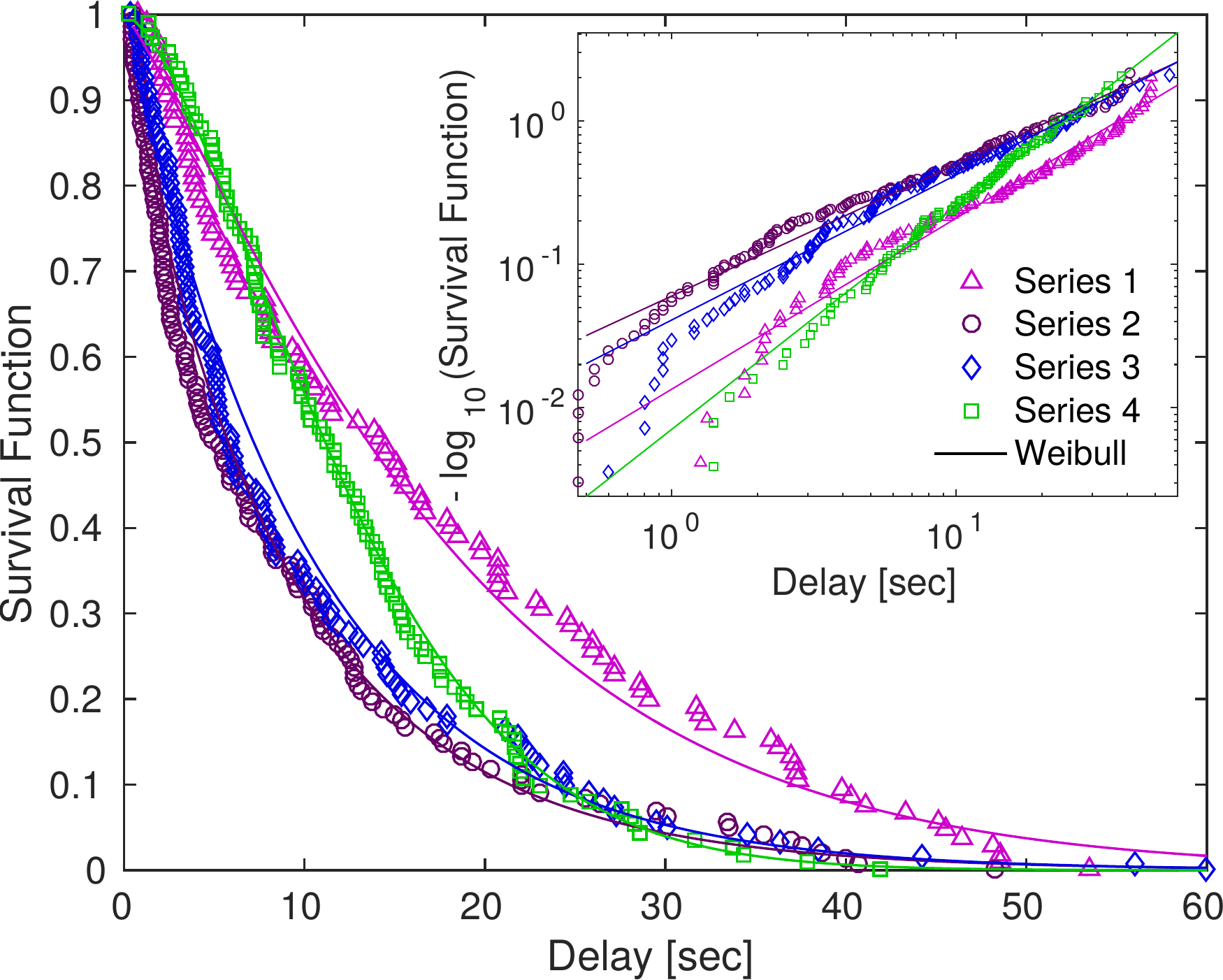}
\caption{Survival function of measured delay times $T$ for the experimental parameters given in Table~\ref{tab:seriespar}, on the subset of data with $n=1$ arms (inset) Weibull plot illustrating linearity in the  large $T$ limit.
Solid curves are numerical fits to a Weibull distribution. Histograms are provided in Supp. Mat. }
\label{fig:distribs}
\end{figure}

For each droplet deposited, we record the delay time $T$. Survival functions for each series are shown in Fig.~\ref{fig:distribs}.
The survival function (or complementary cumulative distribution) is the probability that a droplet survives less than time $T$ before producing fracture(s). We first focus on the set of starbursts with $n=1$.
For all four series, we observe delay times as long as a minute, with the Weibull plot (inset) revealing  all datasets to be highly linear for large $T$. Inevitably, the low-$T$ portion of the histogram contains an excess of data, triggered by non-thermal noise. For example, when we collect data in the presence of additional room noise, we observe that the delay times are systematically reduced from the observations shown here.
Using Matlab's {\tt wblfit()} tool, we fit each waiting time distribution to a Weibull distribution:
$ 
{\cal P}(T) =\left(\frac{\beta}{\tau}\right)\left(\frac{T}{\tau}\right)^{\beta-1} \mathrm{e}^{-( T/\tau )^\beta},
$
as shown in Fig~\ref{fig:distribs}. The corresponding form for the survival function is a stretched exponential $e^{-(T/\tau)^\beta}$; this function takes the form of a straight line on a double-logarithmic plot (see Fig~\ref{fig:distribs}).

The fit parameters ($\tau,\beta$) carry two important interpretations to aid in understanding delayed fracture. The parameter $\tau$ represents a characteristic delay time and the parameter $\beta$ is a shape parameter. For the special case $\beta=1$, the Weibull distribution reduces to an exponential distribution; this corresponds to Poisson-distributed events and constant failure rate. For this case (only), $\tau$ is identical to the mean delay time $\langle T \rangle$.
A shape parameter $\beta > 1$ indicates that the system ages such that failure is more likely the longer the delay.

\begin{figure}
\includegraphics[width=\linewidth]{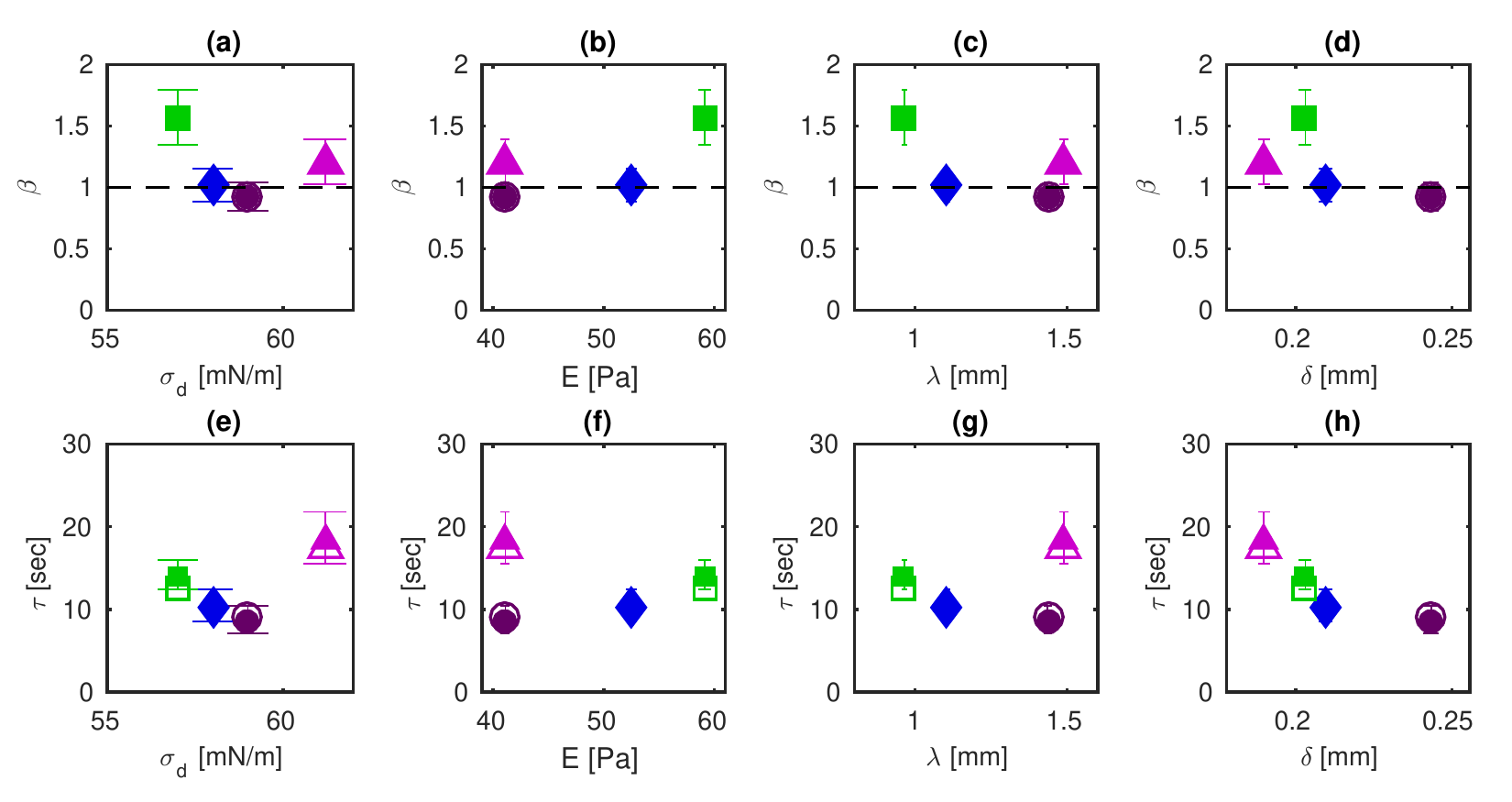}
\caption{Weibull parameters as a function of materials parameters for $n=1$ from the four data series. (a-d) Shape parameter $\beta$ and (e-h) Weibull delay time $\tau$ (solid symbols) with mean delay time $\langle T \rangle$ (open symbols).  Error bars are 95\% confidence intervals on the fit parameters.}
\label{fig:betatau}
\end{figure}

In Fig.~\ref{fig:betatau}, we examine how the material parameters set the failure dynamics, as parameterized by $\beta$ and $\tau$.
For all experiments, $\beta \approx 1$; this indicates that the fracture dynamics are close to a Poisson process. This finding is consistent with what is expected for thermally-activated processes \citep{Bonn-1998-DFI}. Additionally, we observe $\tau \approx \langle T \rangle$ for all four data series, with larger deviations when $\beta$ is further from $1$, as  expected. In two cases (Series 1 \& 4), we observe $\beta >1$; for these values the histograms are non-monotonic (see Supp. Mat.)
This indicates that these gels age in a way that weakens them as a function of time elapsed since loading. Interestingly, Series  1 \& 4 have the least in common when examined in light of their materials parameters (Table~\ref{tab:seriespar}). We additionally observe (see Fig.~\ref{fig:betatau}b) that $\beta$ increases with $E$, and does not monotonically depend on any of the other materials parameters.

The characteristic delay time $\tau$ is observed to decrease for increasing elastocapillary length $\delta$, as shown in Fig.~\ref{fig:betatau}h. One possible interpretation of this result is that larger deformations (larger $\delta$, equivalently larger stresses/strains inside the gels \cite{Bostwick2013,bostwick2014elastocapillary}), lead to shorter delays before fracture.
This decreasing trend is robust when considering  other reasonable values of the gel surface tension in the range  $65 < \sigma_g < 71 $~mN/m (see Supp. Mat.). Over this same range of $\sigma_g$ values, no such systematic trend is observed for the traditional elastocapillary length $\lambda$, based only on $\sigma_d$. This result strongly suggests the surface tension differential $\Delta \sigma$ determines the size of the characteristic force in our experiment. Our observation sheds light on identifying the appropriate traction force boundary condition for partially-wetting substrates in relevant elastocapillary phenomena \citep{roman2010elasto}.

\begin{figure}
\includegraphics[width=0.8\linewidth]{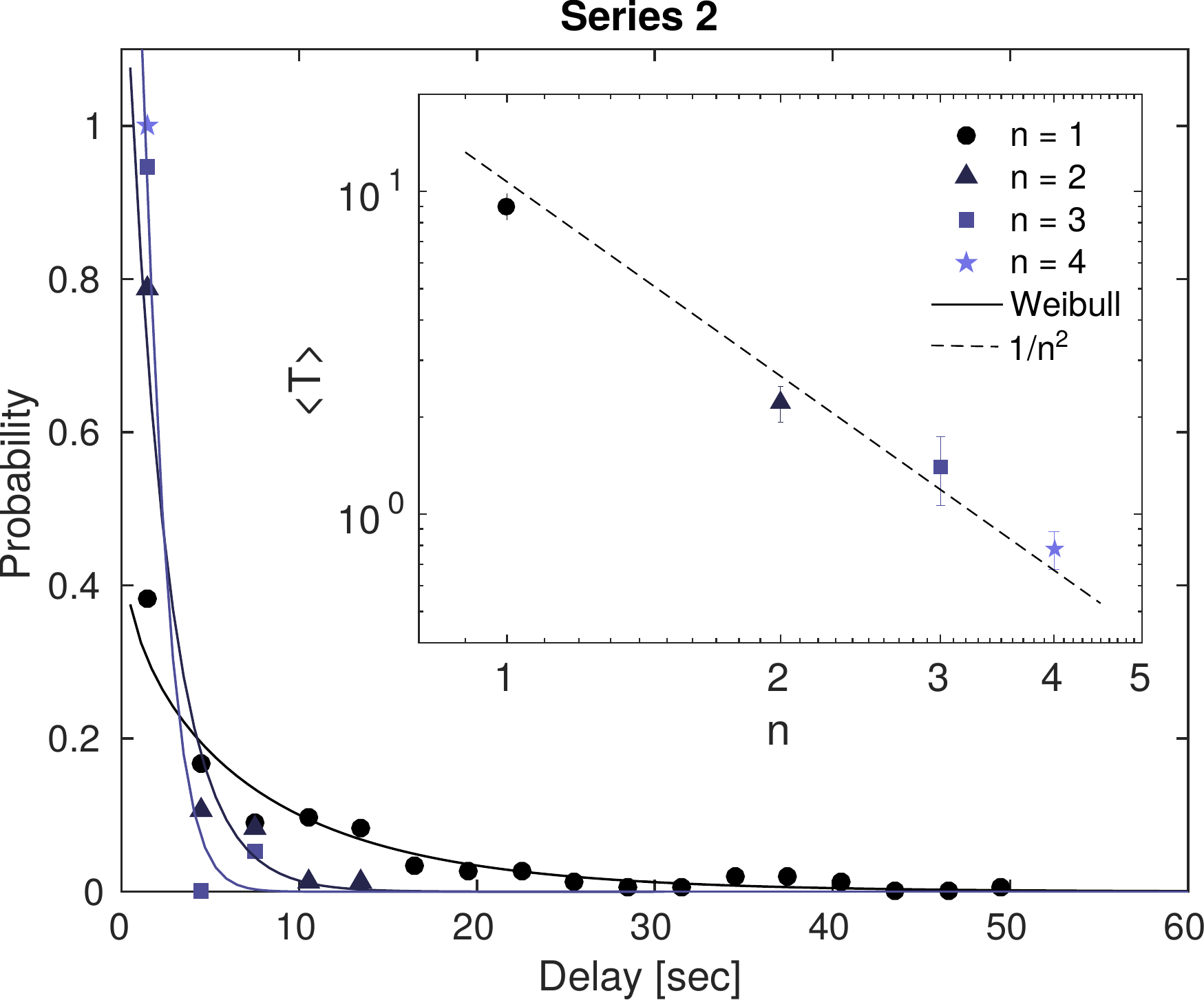}
\caption{Histogram  of measured delay times $T$ for starbursts with a variable number of arms, all from Series 2. (inset) Mean delay time $\langle T \rangle$ as a function of the number of arms. }
\label{fig:multiarmdelay}
\end{figure}

For starbursts with multiple arms, we observe that all of the fractures occur simultaneously, and that delay times are shorter than for the $n=1$ case analyzed above.  This suggests that once one fracture is initiated by a thermal fluctuation, it triggers the simultaneous nucleation of the other fractures. Using the data from Series 2, for which up to $n=4$ fractures were observed, we have sufficient statistics to perform a semi-quantitative investigation.  As shown in Fig.~\ref{fig:multiarmdelay}, we observe that delay times are reduced by a factor proportional to $1/n^2$ for starbursts with multiple arms. For $1 \le n \le 3$, the dynamics are consistent with a Poisson process ($\beta = 1$ within error bars); for $n=4$ there is insufficient data.

\paragraph*{Discussion:}
A liquid droplet deforms an ultrasoft agar substrate to the point of material
failure, resulting in the nucleation of fractures at the contact line. Once nucleated, these capillary fractures propagate
radially outward in a starburst formation, with the mean number of fractures $\langle n \rangle$
controlled by the ratio of the surface tension contrast $\Delta\sigma$ to the elastic modulus $E$ of the gel substrate
\citep{Daniels2007}. We quantify both variations in
the delay time before fractures form, and variations in the number of fractures
within the starburst.

For a given set of experimental parameters (fixed
$E, \sigma_d$, droplet volume $V$), the number of fractures within each starburst
has a well-defined mean, but is not deterministic. Instead, there is a range of
values observed; it is likely that this variability arises from both the
inherent heterogeneity of the gels \citep{Goldbart1989a} and the presence of
multiple unstable deformation modes \citep{Bostwick2013}.
The observed increase in heterogeneity for  small $E$ (small agar
concentration $\phi$) is consistent with models of critical fluctuations on
approach to the gel transition \citep{Goldbart1989a}.

By isolating the case of single-arm starbursts ($n=1$), we infer from the
exponential (or Weibull) distribution of delay times that the
fracture process is thermally-activated. This effect has previously been
observed in systems which are 1000 times stiffer and subject to different
loading conditions \citep{Bonn-1998-DFI}, and this result highlights the universality of delayed fracture dynamics, even for ultrasoft materials.

For a purely thermally-activated process, the delay time distribution ${\cal P}(T)$ would be exponential, with $\tau = \langle T \rangle$ set by the height of the energy barriers in the material. We note that the mean delay time $\langle T \rangle$ is inversely proportional to the thermal nucleation probability ${\cal P}\propto \mathrm{exp}(-\mathcal{E}_{act}/k \mathcal{T})$,
and is thereby a measure of the activation energy $\mathcal{E}_{act}$ \cite{Bonn-1998-DFI,pomeau1992}. Because $\tau$ (or $\langle T \rangle$) are not constant as function of agar concentration $\chi$ (or modulus $E$), our results suggest that the energy associated with the crosslinking of agar is not the only effect. Instead, we observe a trend in which the length scale $\delta = \Delta \sigma / E$ controls the timescale $\tau$. We interpret this finding as highlighting the importance of the {\it differential} in surface tension between the droplet and the gel substrate, rather than just the surface tension of the droplet itself (for which $\lambda$ would have been the key parameter).  Although $\lambda$ traditionally appears in elastocapillary phenomena such as wrinkling, blistering, and stiction \citep{roman2010elasto}, future work is needed to understand whether $\lambda$ or $\delta$ better-describes the degree of deformation, with other material parameters being held constant. Intriguingly, it appears that $E$ is additionally important in controlling the shape parameter of the Weibull distribution, not just the energy barriers, due to its effects on gel aging \cite{Scherer1988,Hodge1995}. Future experiments could use this technique to map out, and disentangle, the effects of aging on the energy barrier landscape in soft materials.

\paragraph*{Acknowledgements:} We would like to thank Michael Shearer and Carlos Ortiz for useful discussions, and Mark Schillaci for crucial upgrades to the experimental protocol.
We are grateful for support from the National Science Foundation under grant
number DMS-0968258.


\newpage 

\begin{samepage}

\begin{figure*}[h]
\centerline{\bf \Large Supplementary Material} 
\bigskip 
\includegraphics[width=0.5\linewidth]{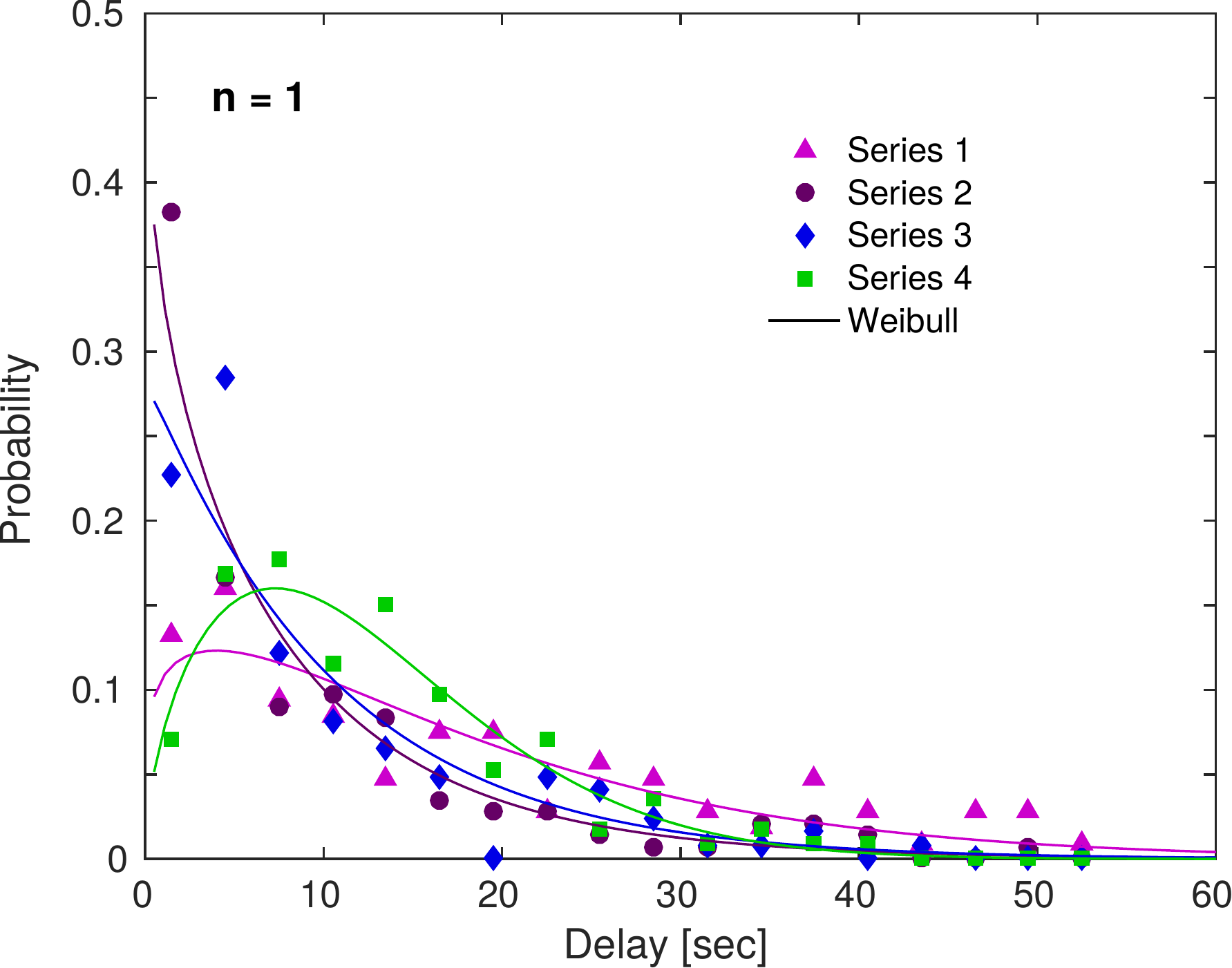}
\caption{Histograms of measured delay times $T$ for the experimental parameters given in Table~I, on the subset of data with $n=1$ arms. Solid curves are numerical fits to a Weibull distribution.}
\end{figure*}

\begin{figure*}[]
\centerline{\hspace{5mm} $\sigma_g = 65$~mN/m  \hfill
$\sigma_g = 66$~mN/m \hfill
$\sigma_g = 67$~mN/m \hfill
$\sigma_g = 68$~mN/m \hfill
$\sigma_g = 69$~mN/m \hfill
$\sigma_g = 70$~mN/m \hspace{5mm} }
\centerline{
\includegraphics[width=2.9cm]{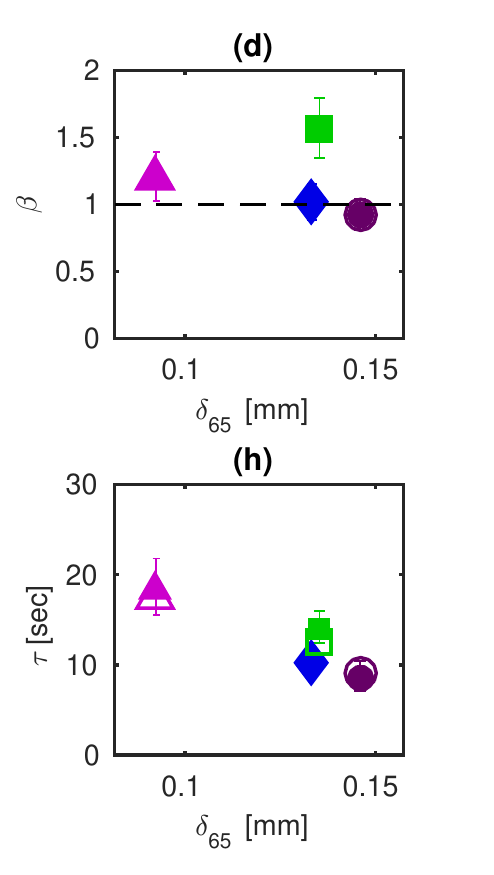}\hfill
\includegraphics[width=2.9cm]{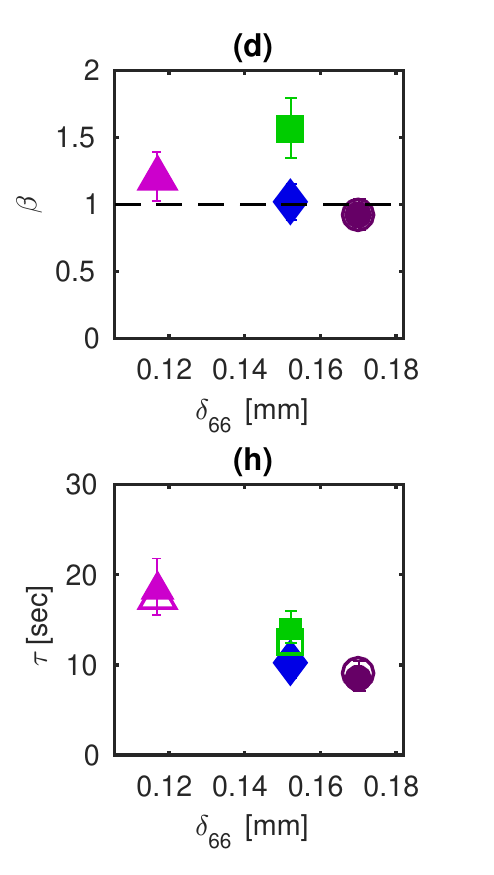}\hfill
\includegraphics[width=2.9cm]{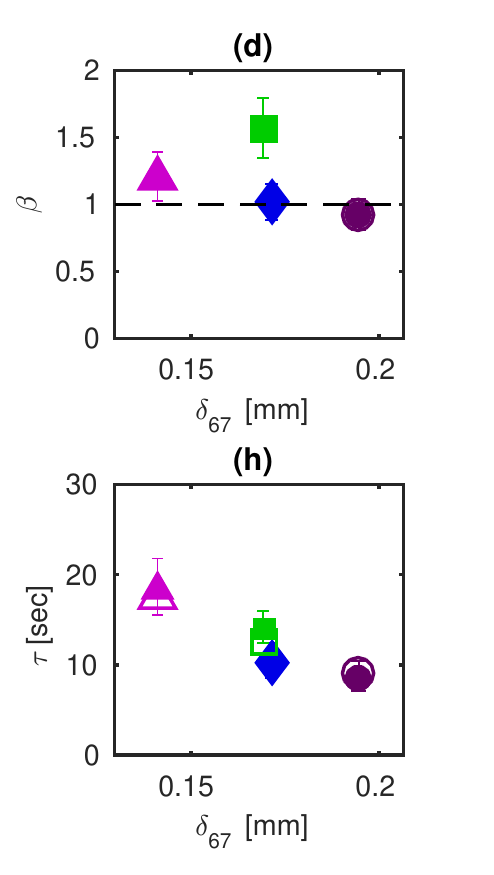}\hfill
\includegraphics[width=2.9cm]{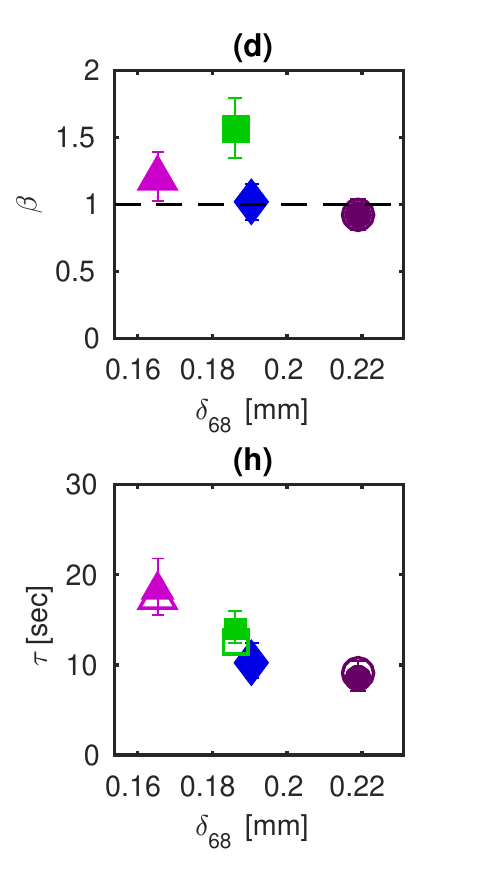}\hfill
\includegraphics[width=2.9cm]{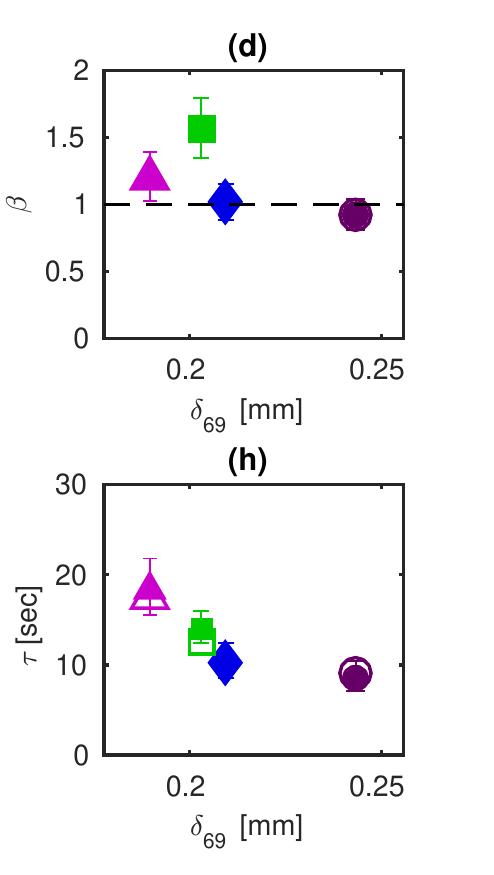}\hfill
\includegraphics[width=2.9cm]{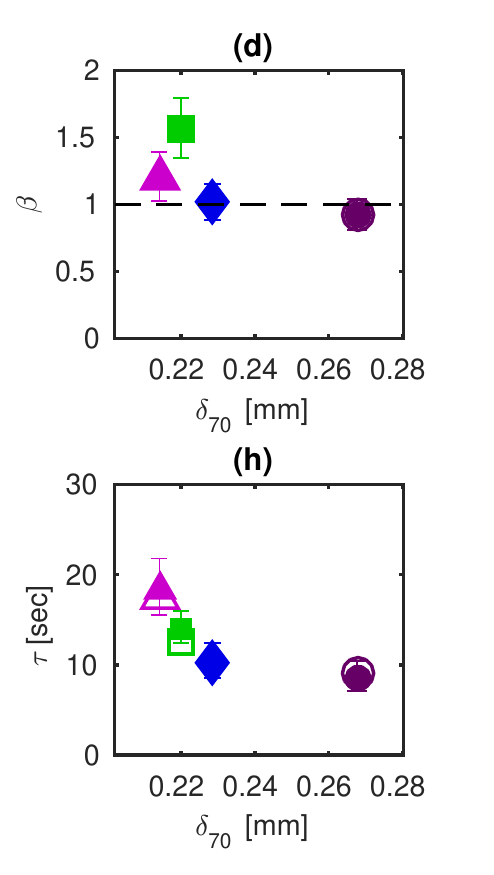}}
\caption{In the main text, we assumed a constant value $\sigma_g = 69$~mN/m for simplicity; this choice affects the value of $\delta = \Delta \sigma / E$ (Eq. 1) and therefore the last column (parts d, h) of Fig. 4. This Figure provides alternative versions of those two  plots, calculated for  different values (65 to 70 mN/m) of $\sigma_g$.}
\end{figure*}

\end{samepage}

\end{document}